\documentclass[letter,traditabstract]{aa} 

\usepackage{savesym}
\usepackage{amsmath}
\savesymbol{iint}
\usepackage{graphicx}
\usepackage{xspace}
\usepackage{booktabs}
\usepackage[varg]{txfonts}
\usepackage{natbib,twoopt}
\usepackage[breaklinks=false, colorlinks=true, citecolor=blue]{hyperref}
\usepackage{epsfig}
\usepackage{txfonts}
\usepackage{amssymb}
\usepackage{hyperref}
\usepackage{color}
\restoresymbol{TXF}{iint}
\bibpunct{(}{)}{;}{a}{}{,} 
 
\newcommandtwoopt{\citeads}[3][][]{\href{http://adsabs.harvard.edu/abs/#3}%
{\citealp[#1][#2]{#3}}}
\newcommandtwoopt{\citepads}[3][][]{\href{http://adsabs.harvard.edu/abs/#3}%
{\citep[#1][#2]{#3}}}
\newcommandtwoopt{\citetads}[3][][]{\href{http://adsabs.harvard.edu/abs/#3}%
{\citet[#1][#2]{#3}}} 
\newcommandtwoopt{\citeyearads}[3][][]%
{\href{http://adsabs.harvard.edu/abs/#3}{\citeyear[#1][#2]{#3}}}

\newcommand{\parm}[1]{\ensuremath{#1}\xspace}

\newcommand{\ptc}{\parm{T_\mathrm{0}}}                    
\newcommand{\pa}{\parm{a}}                                
\newcommand{\pec}{\parm{e}}                               
\newcommand{\pk}{\parm{K}}                                
\newcommand{\pom}{\parm{\omega}}                          
\newcommand{\tperi}{\parm{T_\mathrm{p}}}                  
\newcommand{\gammasys}{\parm{\gamma_\mathrm{sys}}}        
\newcommand{\gammadot}{\parm{\dot{\gamma}}}               

 \def\grcm2{\hbox{\,g\,cm$^{-2}$}}           
 \def\teff{$T_\mathrm{eff}$}                 
 \def\teq{$T_\mathrm{eq}$}                   
 \def\logg{$\log g$}                         
 \def\ms{\hbox{\,m\,s$^{-1}$}}               
 \def\msd{\hbox{\,m\,s$^{-1}$\,d$^{-1}$}}    
 \def\m2s2{\hbox{\,m$^{2}$\,s$^{-2}$}}       
 \def\kms{\hbox{\,km\,s$^{-1}$}}             
 \def\gcm3{\hbox{\,g\,cm$^{-3}$}}            
 \def\vsini{\hbox{$v$\,sin\,$i$}}    
 \def\Msun{\hbox{$M_{\odot}$}}               
 \def\Rsun{\hbox{$R_{\odot}$}}               
 \def\Mjup{\hbox{$\mathrm{M}_\mathrm{Jup}$}} 
 \def\Rjup{\hbox{$\mathrm{R}_\mathrm{Jup}$}} 
 \def\mp{{\emph M}$_\mathrm{p}$}             
 \def\rp{{\emph R}$_\mathrm{p}$}             
 \def\kepler{\emph{Kepler}}                  

\newcommand{\pilar}{Pilar\,Monta\~n\'es-Rodr\'{\i}guez}
\begin{document} 

 \title{Kepler-432 b: a massive warm Jupiter in a 52-day eccentric orbit transiting a giant star\thanks{Based on observations collected at 
   the German-Spanish Astronomical Center, Calar Alto, jointly operated by the Max-Planck-Institut f\"{u}r Astronomie (Heidelberg) 
   and the Instituto de Astrof\'isica de Andaluc\'ia (IAA-CSIC, Granada).}\fnmsep\thanks{Based on observations obtained with the 
   Nordic Optical Telescope, operated on the island of La Palma jointly by Denmark, Finland, Iceland, Norway, and Sweden, in the 
   Spanish Observatorio del Roque de los Muchachos of the Instituto de Astrof\'isica de Canarias.}}

   \author{Mauricio\,Ortiz\inst{1},
           Davide\,Gandolfi\inst{1}, 
           Sabine\,Reffert\inst{1},   
           Andreas\,Quirrenbach\inst{1}
           \and
           Hans\,J.\,Deeg\inst{2,3}
           \and
           Raine\,Karjalainen\inst{4}
           \and 
           \pilar\inst{2,3}
           \and 
           David\,Nespral\inst{2,3}
           \and 
           Grzegorz\,Nowak\inst{2,3}
           \and 
           Yeisson\,Osorio\inst{5}
           \and 
           Enric\,Palle\inst{2,3}
           }

   \institute{Landessternwarte, Zentrum f\"{u}r Astronomie der Universit\"{a}t Heidelberg, K\"{o}nigstuhl 12, 69117 Heidelberg, Germany\\
              \email{mortiz@lsw.uni-heidelberg.de}                                  
   \and
               Instituto de Astrof\'isica de Canarias, C. V\'ia L\'actea S/N, E-38205 La Laguna, Tenerife, Spain
    \and
                 Departamento de Astrof\'isica, Universidad de La Laguna, E-38200 La Laguna, Tenerife, Spain
        \and 
                Isaac Newton Group of Telescopes, Apartado de Correos 321, E-38700 Santa Cruz de Palma, Spain
         \and 
                Nordic Optical Telescope, Apartado 474, 38700 Santa Cruz de La Palma, Spain
                 }
   \date{Received 11 October 2014 / Accepted 28 November 2014}

 
  \abstract{We study the \textit{Kepler} object Kepler-432, an evolved star ascending the red giant branch. By deriving precise radial 
  velocities from multi-epoch high-resolution spectra of Kepler-432 taken with the CAFE spectrograph at the 2.2m telescope of Calar 
  Alto Observatory and the FIES spectrograph at the Nordic Optical Telescope of Roque de Los Muchachos Observatory, we confirm the 
  planetary nature of the object Kepler-432 b, which has a transit period of 52 days. We find a planetary mass of \mp=$5.84\pm0.05$~\Mjup\ 
  and a high eccentricity of $e$=$0.478\pm0.004$. With a semi-major axis of $a$=$0.303\pm0.007$~AU, Kepler-432 b is the first 
  \textit{bona fide} warm Jupiter detected to transit a giant star. We also find a radial velocity linear trend of 
  $\dot{\gamma}$=$0.44\pm0.04$~\msd, which suggests the presence of a third object in the system. Current models of planetary evolution in 
  the post-main-sequence phase predict that Kepler-432 b will be most likely engulfed by its host star before the latter reaches the tip 
  of the red giant branch.}

\titlerunning{The transiting warm Jupiter Kepler-432 b}
\authorrunning{M. Ortiz, D. Gandolfi, S. Reffert, A. Quirrenbach et al.}

   \maketitle
%

\section{Introduction}
The number of Jupiter-like planets found to orbit evolved stars has constantly been growing in recent years. This includes planets around 
subgiant \citep{Johnson2010,Johnson2011} and giant stars \citep{Gettel2012,Sato2013,Trifonov2014}. These discoveries have provided evidence 
that the gas-giant planet population around evolved stars possesses different orbital properties than the population orbiting main-sequence (MS) 
stars \citep[e.g.,][]{Jones2014a}. The most notable trend is the apparent lack of close-in Jupiter-like planets orbiting giant or subgiant 
stars, although these objects are easily found around many MS stars by Doppler surveys \citep[e.g.,][]{Jones2013}. 
Specifically, there seems to be a lack of planets around giant stars with semi-major axis $a<0.5$~AU. Exceptions to this apparent trend 
are the recently discovered planet \object{HIP67851 b} \citep{Jones2014b} and \object{Kepler-91 b}, the only hot Jupiter known to transit a 
giant star \citep{Lillo2014}.

There are two different mechanisms that have been proposed to explain the paucity of close-in Jupiter-like planets around giant stars. 
The first one states that, as a result of stellar evolution, the inner planets are tidally engulfed by their host stars as the outer planets move 
farther out \citep{Kunitomo2011,Adamow2012,Schlaufman2013}. The second one suggests that although giant planets may form around 
intermediate-mass stars, they do not migrate inwards, owing to the short dissipation time-scale of protoplanetary disks 
\citep{Kretke2009,Currie2009}. By searching specifically for close-in Jupiter-like planets around giant stars, we can help to 
place constraints on the theoretical models that try to explain these observations and, possibly, also learn something about the evolution 
of planetary systems after the host star leaves the MS.

In this letter, we confirm and characterize Kepler-432 b, a massive warm gas-giant planet orbiting a star ascending 
the red giant branch (RGB). 


\section{High-resolution spectroscopic follow-up}
\label{Spectroscopy}
The radial velocity (RV) follow-up of Kepler-432 was carried out between June and October 2014 using the Calar Alto Fiber-fed \'Echelle 
spectrograph \citep[CAFE;][]{Aceituno2013} -- mounted at the 2.2m telescope of Calar Alto Observatory (Almer\'ia, Spain) -- and the 
FIbre-fed \'Echelle Spectrograph \citep[FIES;][]{Telting2014} -- mounted at the 2.56m Nordic Optical Telescope of Roque 
de los Muchachos Observatory (La Palma, Spain). We acquired 11 RVs with CAFE (R$\sim$62000), and 16 with FIES (R$\sim$67000) at 
different epochs. To remove cosmic-ray hits, three consecutive exposures were usually taken per epoch observation. Following the observing 
strategy described in \citet{Aceituno2013} and \citet{Buchhave2010}, we traced the RV drift of CAFE and FIES by acquiring long-exposed 
(T$_\mathrm{exp}$=60--80 sec) ThAr spectra immediately before and after each epoch observation. The data were reduced using IRAF and IDL 
standard routines, which include bias subtraction, flat fielding, order tracing and extraction, and wavelength calibration. Radial 
velocities were derived via multi-order cross-correlation with the RV standard stars \object{HD\,182572} (CAFE) 
and \object{HR\,5777} (FIES).

\begin{figure}
\centering
\includegraphics[width=0.5\textwidth]{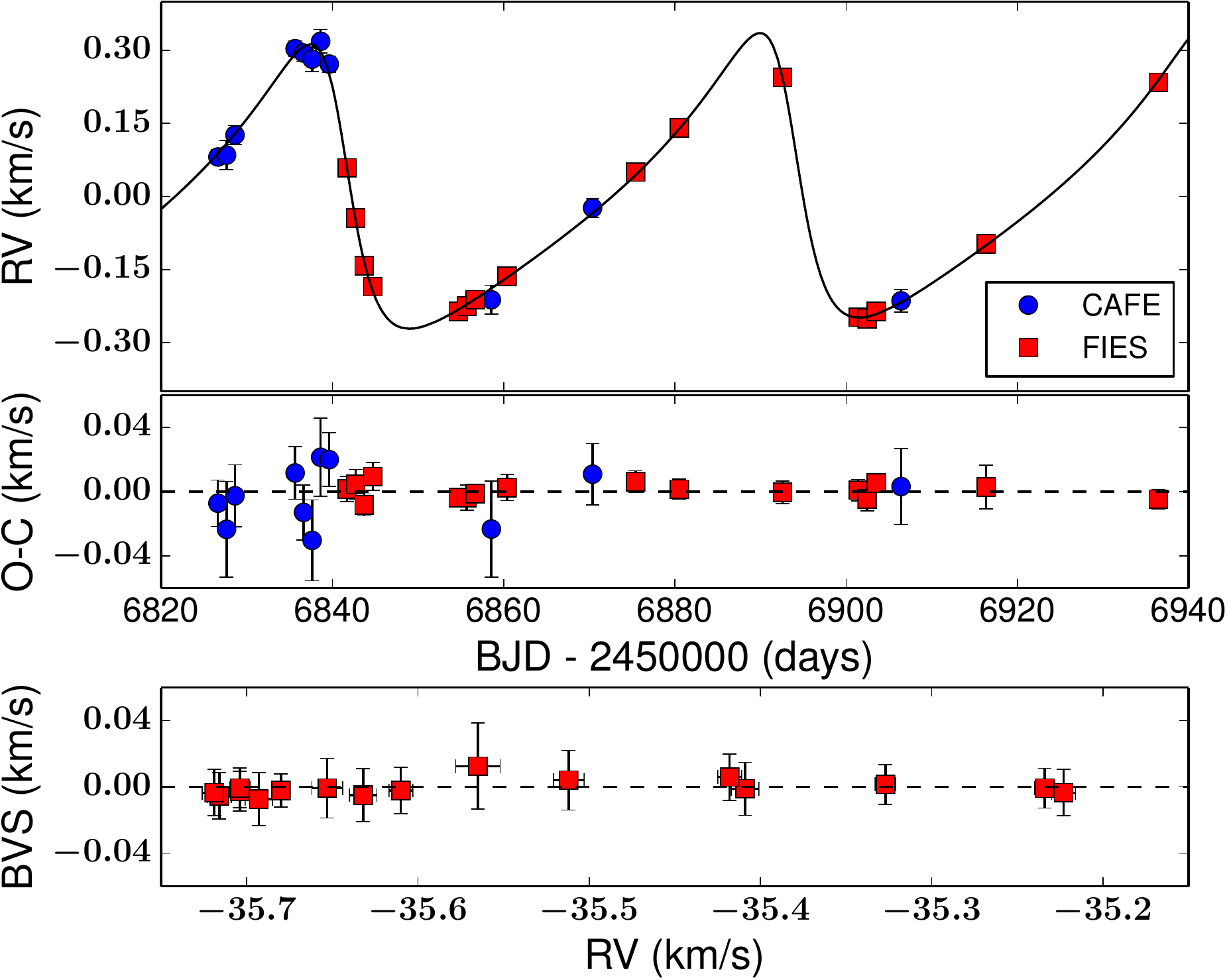}
\caption{Radial velocity measurements of Kepler-432. \textit{Upper panel}: CAFE (blue circles) and FIES (red squares) RVs, and Keplerian fit 
to the data (black solid line) -- including the linear RV trend. \textit{Middle panel}: RV residuals. The rms is $\sim$17 m/s and 
$\sim$5 m/s for the CAFE and FIES data, respectively. The observed rms of the FIES RVs is consistent with the expected 
value of $\sim6$ m/s for a star with $\log g = 3.35$ dex \citep{Hekker2008, Nowak2013}. Quadratically adding a jitter of 6 m/s to 
our formal RV measurement errors does not change the derived orbital parameters significantly (< 1$\sigma$). Additionally, a fit 
to the FIES data alone yields consistent results within 1$\sigma$. \textit{Lower panel}: Bisector velocity spans (BVS) of the FIES CCF 
versus RVs, assuming that the error bars of the former are twice those of the latter.}
\label{fig:1}
\end{figure}

The CAFE and FIES RVs are listed in Table \ref{RV-Table} -- along with their uncertainties, total exposure times, signal-to-noise (S/N) 
ratios per pixel at 5500~\AA, and cross-correlation function (CCF) bisector spans -- and are plotted in Fig. \ref{fig:1} together with 
the Keplerian fit to the data (upper panel) and residuals to the fit (middle panel). The lower panel of Fig. \ref{fig:1} shows the FIES CCF 
bisector spans plotted against the RV measurements, assuming that the error bars of the former are twice those of the latter. We 
followed the method described in \citet{Loyd2014} to account for the uncertainties of our measurements and found a $\sim$50\% probability 
that an uncorrelated set of points (null hypothesis) can reproduce the data. The lack of a significant correlation between the CCF bisector 
spans and the RVs indicates that the Doppler shifts observed in Kepler-432 are most likely induced by the orbital motion of a planet and 
not by a blended eclipsing binary or stellar activity \citep[see, e.g.,][]{Queloz2001}. Moreover, the analysis of the centroid motion during 
the transit reported in the \kepler\ data validation report excludes, at the 3-$\sigma$ level, any scenario in which the transit signal 
is caused by a contaminating eclipsing binary at a distance of more than 0.4$\arcsec$ from the source.

\section{Results}
\subsection{Stellar properties}
We determined the spectral parameters of Kepler-432 by fitting the co-added FIES spectrum (S/N$\sim$145 per pixel at 5500\,\AA) to a grid of 
synthetic spectra calculated with the SPECTRUM code \citep{Gray1994} using ATLAS9 models \citep{Castelli2004}. Microturbulent 
($v_ {\mathrm{micro}}$) and macroturbulent ($v_ {\mathrm{macro}}$) velocities were derived following \citet{Hekker2007}. Stellar mass and 
radius were determined via the asteroseismic scaling relations given by \citet{White2011}, using our estimate for the effective temperature 
along with the large frequency separation and the frequency at which oscillations have the maximum power, as derived by \citet{Huber2013}. 
We estimated the stellar age using theoretical isochrones from \citet{Bressan2012}. Distance and interstellar extinction were calculated 
following the method described in \citet{Gandolfi2008}. The derived stellar parameters are listed in Table \ref{Star-parm}.

\begin{table}[t]
\centering
\caption{Stellar parameters of Kepler-432.}
\label{Star-parm}
\scalebox{0.9}{
\begin{tabular}{lc}
            \hline
            \hline
            \noalign{\smallskip}
            Parameter      &  \text{Value} \\
            \noalign{\smallskip}
            \hline
            \noalign{\smallskip}
            Effective temperature \teff\ (K)                         & $5020\pm60$              \\
            Spectroscopic surface gravity \logg\ (log$_{10}$ \grcm2) & $3.35\pm0.07$            \\
            Metallicity [M/H] (dex)                                  & $-0.02\pm0.06$           \\
            Microturbulent velocity $v_ {\mathrm{micro}}$ (\kms)     & $1.3\pm0.3$              \\
            Macroturbulent velocity $v_ {\mathrm{macro}}$ (\kms)     & $3.5\pm0.5$              \\
            Projected rotational velocity \vsini\ (\kms)             & $\leq 1$                 \\
            Stellar mass $M_\star$ (\Msun)                           & $1.35\pm0.10$            \\
            Stellar radius $R_\star$ (\Rsun)                         & $4.15\pm0.12$            \\
            Age (Gyr)                                                & $3.6^{+1.0}_{-0.6}$      \\
            Distance (pc)                                            & $874\pm30$               \\
            Interstellar extinction $A_\mathrm{V}$ (mag)             & $0.26\pm0.04$            \\
            Spectral type                                            & \text{K}2\,\text{III}    \\
            \noalign{\smallskip}
            \hline
\end{tabular}
}
\end{table}

\subsection{Orbit and planet parameters}

We fitted a Keplerian orbit to the RV data using the IDL code RVLIN \citep{Wright2009}. Uncertainties of the derived parameters were 
estimated using the bootstrap procedure described in \citet{Wang2012}. Orbital period $P_\mathrm{orb}$ and mid-transit epoch \ptc\ were 
fixed to the values reported in the Kepler objects of interest (KOI) database.

We fitted for the eccentricity \pec, argument of periastron \pom, radial velocity semi-amplitude \pk, periastron time \tperi, systemic RV 
\gammasys, fixed zero point RV offset between CAFE and FIES data-sets, and RV linear trend \gammadot. Fixing \gammadot$=0$ leads to a poor 
fit to the data ($\chi_\mathrm{red}^2=3.5$), with a systematic offset from the RV measurements. The FIES RV residuals -- that is, those with 
the smaller error bars -- show a significant correlation with time if no trend is considered, the correlation coefficient 
being 0.86 with a false-alarm probability lower than 0.9\%. Therefore, we consider the trend in the RVs to be real and obtain a value 
of $\gammadot$=$0.44\pm0.04$ \msd. This is significant at the 11-$\sigma$ level and most likely due to an additional companion in the 
system, whose nature remains to be established. We report the best-fit orbital parameters in Table \ref{Planet-parm}.

%
\section{Discussion}
\subsection{Internal structure and equilibrium temperature}

\begin{table}[t]
\centering
\caption[]{Orbital parameters of Kepler-432 b.}
\label{Planet-parm}
\scalebox{0.9}{
\begin{tabular}{lc}
      
            \hline
            \hline
            \noalign{\smallskip}
            Parameter      &  \text{Value} \\
            \noalign{\smallskip}
            \hline
            \noalign{\smallskip}
            RV semi-amplitude \pk (\ms)                 & $294.6\pm2.1$      \\
            Eccentricity \pec                           & $0.478\pm0.004$    \\
            Argument of periastron \pom (deg)           & $68.4\pm0.7$       \\
            Periastron time \tperi (BJD-2\,450\,000)    & $6841.06\pm0.03$   \\
            Systemic velocity \gammasys (\kms)          & $-33\pm0.3$    \\
            CAFE-FIES offset velocity (\ms)             & $634\pm5$          \\
            RV linear trend \gammadot (\msd)            & $0.44\pm0.04$      \\
            Planet mass \mp~(\Mjup)                     & $5.84\pm0.05$      \\
            Semi-major axis \pa (AU)                    & $0.303\pm0.007$    \\
            \noalign{\smallskip}
            \hline
\end{tabular}

}
\tablefoot{Period and mid-transit epoch fixed to $P_\mathrm{orb}$=$52.5010768$ days and $\ptc$=$5004.519$ (BJD-2\,450\,000). 
To derive the true planet mass, we use an orbital inclination value of $i$=$89.95^{\circ}$ from the KOI database.}      
\end{table}

\begin{figure}[t]
\centering
\includegraphics[width=0.49\textwidth]{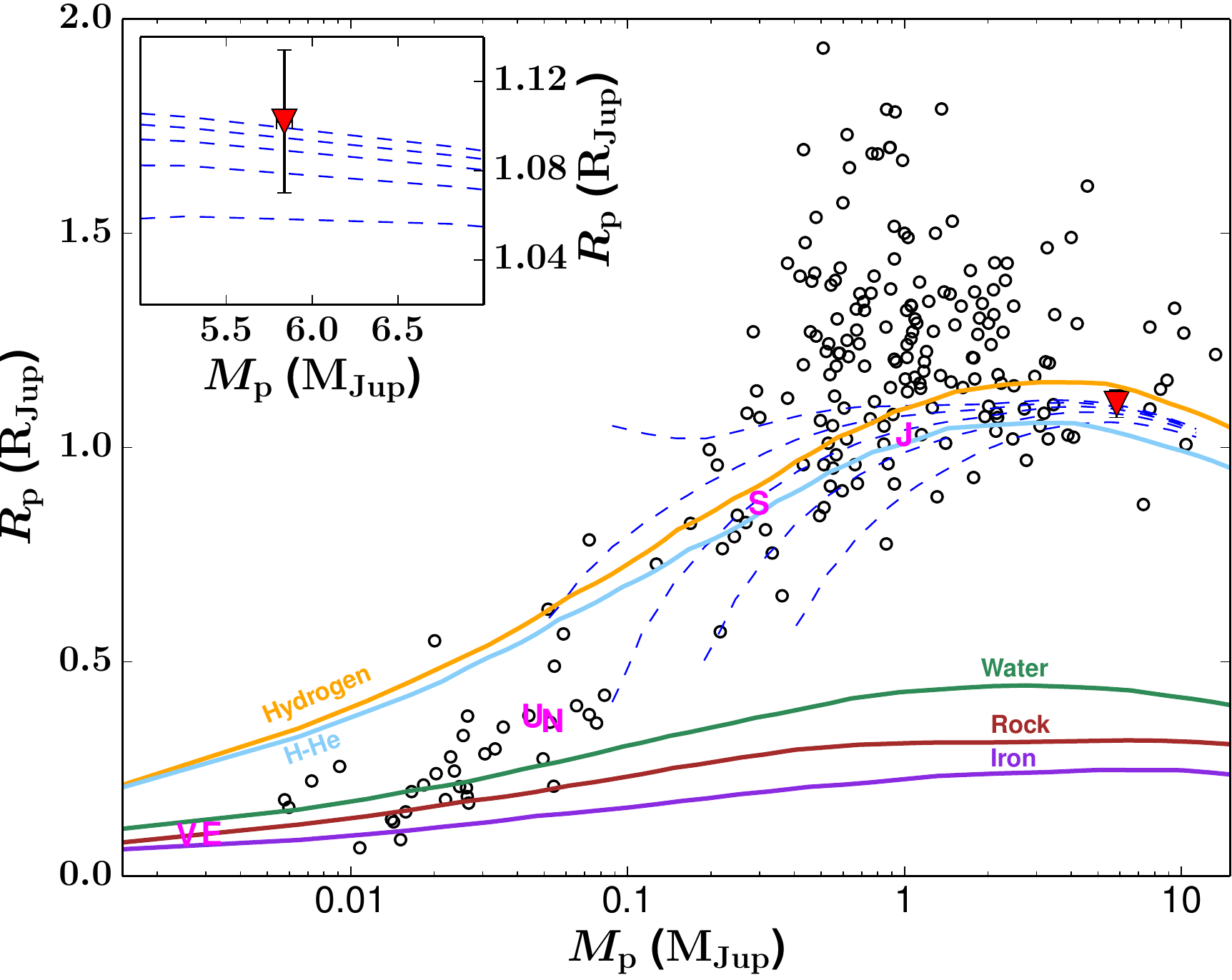}
\caption{Radius and masses of the known transiting exoplanets (black empty circles). The position of Kepler-432 b is marked by the red triangle. The \citet{Fortney2007} isochrones for planet core masses of 0, 10, 25, 50, and 100 $M_{\oplus}$ -- interpolated to the insolation and age of Kepler-432 b -- are overplotted with dashed lines from top to bottom. The upper left inset is a zoom around Kepler-432 b. Also shown are models for planets 
of different compositions derived by \citet{Seager2007}. Solar system planets are marked with magenta letters. We note 
that Kepler-432 b falls in a region with a lack of planets between $\sim$4.5 and $\sim$7~\Mjup.}
\label{fig:2}
\end{figure}

Kepler-432 b is the first \textit{bona fide} confirmed transiting warm Jupiter found to orbit a red giant star. About 70\% of the known transiting giant 
planets have densities in the range 0.35--1.20\,g\,cm$^{-3}$ and masses between 0.3 and 3\,\Mjup, with a peak around 1\,\Mjup\ 
(Fig. \ref{fig:2}). With \mp=$5.84\pm0.05$~\Mjup\ and $\rho_\mathrm{p}$=$5.4\pm0.5$~g\,cm$^{-3}$, Kepler-432 b is one of the most dense and 
massive gas-giant planets known so far. The mass of Kepler-432 b agrees with the general trend found by Doppler surveys, that is, planets around 
giant stars tend to be more massive (3-10~\Mjup) than planets orbiting solar-like stars \citep[see e.g.,][]{Dollinger2009,Reffert2015}. In 
fact, around 96\% of the known planets orbiting solar-type MS stars have masses lower than $5$~\Mjup.

We investigated the internal structure of Kepler-432 b using the models from \citet{Fortney2007}, which couple planetary evolution to stellar 
irradiation for H-He-rich planets (dashed blue lines in Fig. \ref{fig:2}). The planet radius of \rp=$1.102\pm0.032$~\Rjup~is consistent within 2-$\sigma$ with theoretical values for giant planets with core masses of  $\lesssim$~100~M$_{\oplus}$. This implies that 
the planet core accounts most likely for 6\% or less of the total mass, that is, similar to the fractional core mass of 
Jupiter \citep{Saumon2004}. We note that the solar-like metallicity of the host star [Fe/H]=$-0.02\pm0.06$\,dex supports the low-fractional core 
mass scenario for Kepler-432 b.

As a consequence of the high eccentricity of the orbit ($e$=$0.478\pm0.004$), the planet is at nearly $\sim$0.16~AU ($\sim$8\,$R_{\star}$) 
from its host star during periastron, receiving a flux of $F_\mathrm{per}$=$(6.1\pm0.6)\times10^{8}$\,erg\,s$^{-1}$\,cm$^{-2}$. At 
apastron, Kepler-432 b travels as far out as $\sim$0.45~AU ($\sim$23\,$R_{\star}$), receiving an incoming radiation of 
$F_\mathrm{apo}$=$(7.6\pm0.8)\times10^{7}$\,erg\,s$^{-1}$\,cm$^{-2}$. The time-averaged incident flux on Kepler-432 b is 
$\langle\,F\,\rangle$=$(1.66\pm0.17)\times10^{8}$\,erg\,s$^{-1}$\,cm$^{-2}$.

Assuming a black-body stellar emission and planetary heat redistribution factor $f$ between 0.25 (instantaneous energy redistribution into 
the planet atmosphere) and 0.67 \citep[instantaneous energy reradiation to space; see][]{Lopez2007}, we derive an average equilibrium 
temperature of \teq=$943\pm20$\,K for a bond albedo of 0.27 \citep[based on][]{Kane2010}. Following the planet 
classification in \citet{Sudarsky2000}, Kepler-432 b would be a member of the class IV planets with temperatures in the range 
900\,$<$\,\teq\,$<$\,1500 K, for which a tropospheric silicate layer is expected to exist. However, we note that eccentric orbits can lead 
to significant changes in the atmospheric compositions, owing to the large variation of the incident stellar flux \citep{Sudarsky2005}. 
For Kepler-432 b, we expect temperature differences of $\sim 500$ K between periastron and apastron.

\subsection{Kepler-432 b: a close-in eccentric planet}
Among planets orbiting giant stars, Kepler-432 b is very peculiar both in terms of eccentricity and orbital period, as it occupies scarcely 
populated regions of the $P_\mathrm{orb}$ versus M$_*$ and $a$ versus $e$ diagrams (Fig. \ref{fig:3}). While planets with orbital periods 
between 1 and 10$^{4}$~days are common around MS stars, there is a clear lack of short-period planets around giant stars, and Kepler-432 b 
is one of the few inhabitants of the region with $P_\mathrm{orb}<100$~days.

The value of $e$=$0.478\pm0.004$ for the eccentricity is among the highest for planets orbiting giant stars. Most planets around giant 
stars tend to have low eccentricity (63\% have $e<0.2$), with a median of $e=0.15$, whereas planets around MS stars tend to be more 
eccentric. If we consider objects with $a\gtrsim0.5$~AU -- where most of the planets around giant stars are found -- only 39\% of 
planets orbiting solar-type stars exhibit $e<0.2$, and a K-S test gives a probability of 0.1\% that the eccentricity of planets around 
giant and MS stars is drawn from the same distribution.

\onltab{
\begin{table*}
\caption{CAFE and FIES radial velocity measurements of Kepler-432. The total exposure time,  S/N ratio per pixel at 5500 \AA, and CCF bisector spans are listed in the last three columns.}
\label{RV-Table}
\centering
\begin{tabular}{c c c c c r}
\hline
\hline
\noalign{\smallskip}
BJD              &   RV    & $\sigma_{\mathrm RV}$ & Exp. Time &  S/N/pixel   &   \multicolumn{1}{c}{BVS}   \\
($-$ 2\,450\,000)&   \kms  &    \kms               &    sec    &  @5500\,\AA  &   \multicolumn{1}{c}{\kms}  \\
\noalign{\smallskip}
\hline
\noalign{\smallskip}
CAFE & & & & \\
\noalign{\smallskip}
6826.63307  & $-$36.018 & 0.014 & 5400 & 25 & \multicolumn{1}{c}{--}     \\
6827.64045  & $-$36.017 & 0.029 & 5400 & 16 & \multicolumn{1}{c}{--}     \\ 
6828.62402  & $-$35.976 & 0.019 & 5400 & 22 & \multicolumn{1}{c}{--}     \\
6835.64888  & $-$35.799 & 0.016 & 5400 & 26 & \multicolumn{1}{c}{--}     \\
6836.63077  & $-$35.808 & 0.017 & 5400 & 22 & \multicolumn{1}{c}{--}     \\
6837.63356  & $-$35.820 & 0.025 & 5400 & 16 & \multicolumn{1}{c}{--}     \\
6838.63076  & $-$35.783 & 0.024 & 5400 & 20 & \multicolumn{1}{c}{--}     \\
6839.63282  & $-$35.830 & 0.017 & 5400 & 24 & \multicolumn{1}{c}{--}     \\
6858.56617  & $-$36.314 & 0.029 & 5400 & 18 & \multicolumn{1}{c}{--}     \\
6870.39070  & $-$36.125 & 0.019 & 5400 & 19 & \multicolumn{1}{c}{--}     \\
6906.41528  & $-$36.316 & 0.023 & 4500 & 21 & \multicolumn{1}{c}{--}     \\

\hline
\noalign{\smallskip}
FIES & & & & \\
\noalign{\smallskip}
6841.71887 & $-$35.409 & 0.008 & 1800 & 31 & $-$0.001 \\
6842.71994 & $-$35.512 & 0.009 & 1800 & 30 &    0.004 \\
6843.71619 & $-$35.610 & 0.007 & 1800 & 33 & $-$0.002 \\
6844.72316 & $-$35.653 & 0.009 & 1680 & 29 & $-$0.001 \\
6854.70204 & $-$35.704 & 0.006 & 2400 & 41 & $-$0.003 \\
6855.69674 & $-$35.693 & 0.008 & 2700 & 44 & $-$0.007 \\
6856.68809 & $-$35.680 & 0.005 & 2700 & 41 & $-$0.002 \\
6860.40078 & $-$35.632 & 0.008 & 1800 & 34 & $-$0.005 \\
6875.41274 & $-$35.418 & 0.007 & 2400 & 36 &    0.006 \\
6880.52282 & $-$35.327 & 0.006 & 2400 & 44 &    0.002 \\
6892.55722 & $-$35.223 & 0.007 & 2400 & 35 & $-$0.004 \\
6901.41459 & $-$35.716 & 0.007 & 2400 & 36 & $-$0.006 \\
6902.46355 & $-$35.719 & 0.007 & 2400 & 31 & $-$0.004 \\
6903.52509 & $-$35.704 & 0.006 & 2400 & 38 & $-$0.001 \\
6916.36055 & $-$35.565 & 0.013 & 1800 & 24 &    0.012 \\
6936.46518 & $-$35.234 & 0.006 & 2400 & 44 & $-$0.001 \\
\noalign{\smallskip}
\hline
\end{tabular} 
\end{table*}
}

Because Kepler-432 b is dynamically young \citep[circularization time scale $\tau_\mathrm{circ}\sim150$ Gyr; see][]{Jackson2008},  the 
non-zero eccentricity of the planet might be a tracer of its migration history. In this context, it is expected that some kind of 
high-eccentricity migration (HEM) mechanism might have operated to excite the eccentricity of Kepler-432 b to its current value 
\citep[see][and references therein]{Socrates2012}. Within this scenario, it has been proposed that warm Jupiters can form via tidal 
dissipation at the high-$e$ stage during Kozai-Lidov oscillations \citep{Wu2011,Dong2014}. Moreover, \citet{Dong2014} suggested that 
these planets need close companions for HEM to occur. They calculated upper limits for the perturber separation $b_\mathrm{per}$ to allow 
an efficient tidal dissipation of the orbit. Assuming a perturber mass of between 1 and 10\,\Mjup, we find an upper limit of 
$b_\mathrm{per}$ =  6\,--\,13~AU for an additional object in the Kepler-432 system.

Following \citet{Montet2014}, for example, the detected RV acceleration $\dot{\gamma}=0.44$ \msd\ requires a perturbing object 
in a circular orbit at $a\sim1$ and $a\sim3$~AU for masses of 1 and 10~\Mjup, respectively, consistent with the HEM scenario. We consider it 
unlikely that spot-induced variability might be the source of the RV drift because no emission in the core of the Ca H and K lines is detected in the FIES co-added spectrum. Furthermore, \citet{McQuillan2013} found no significant modulation in the \textit{Kepler} light 
curve of Kepler-432, which suggests that this is a magnetically quiet star. A second companion in the system might thus account for the observed RV drift. Additional observations are needed to better assess the nature of this trend.

\subsection{Post-main-sequence evolution}
\begin{figure}
\centering
\includegraphics[width=0.49\textwidth]{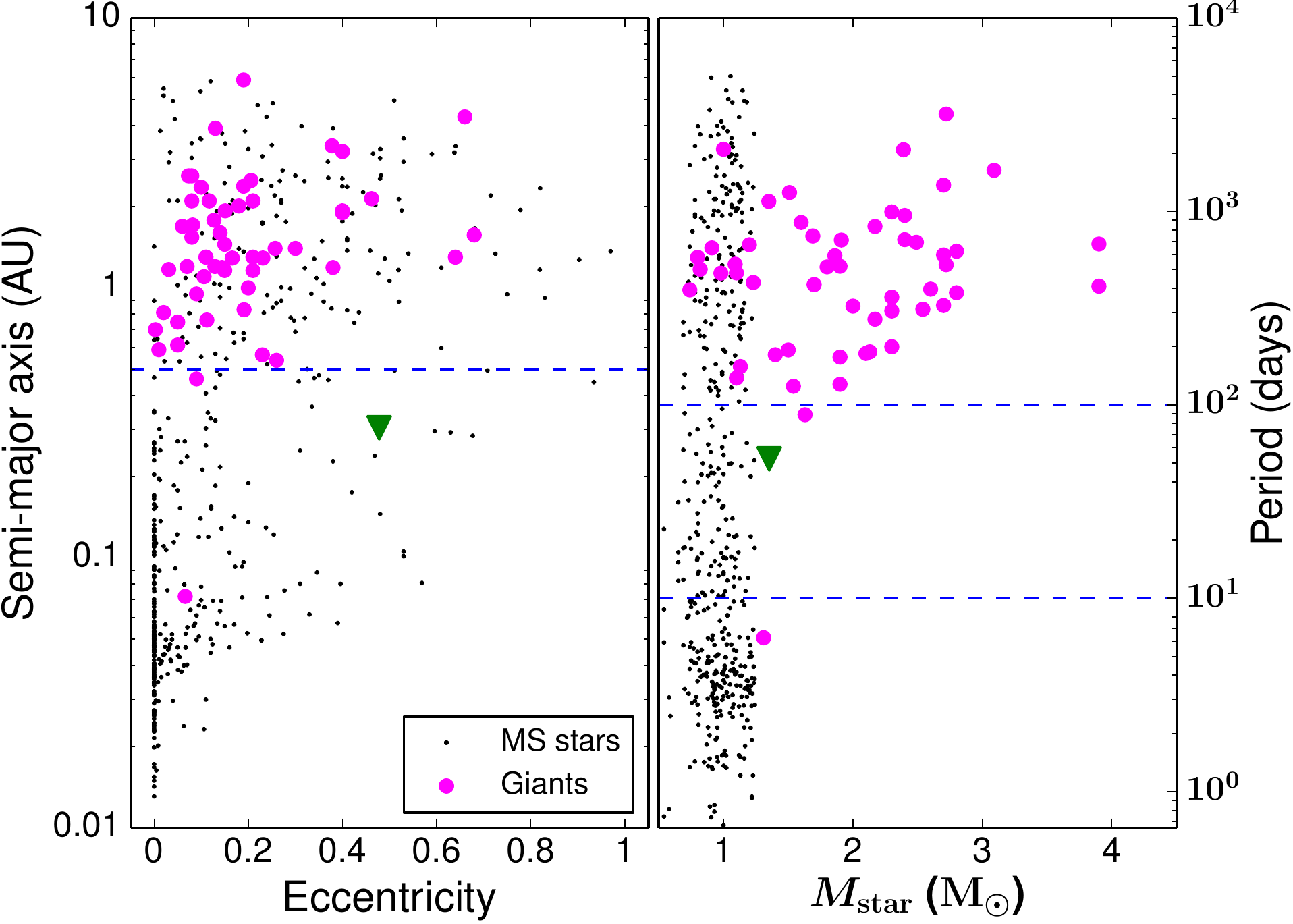}
\caption{\textit{Left panel:} Eccentricity and semi-major axis of the extrasolar planets discovered around MS stars (black dots) and 
giant stars (magenta circles). The dashed line shows the value of $a=0.5$~AU. \textit{Right panel:} Orbital period versus stellar mass. 
The dashed lines represent the region of $10 \leq P_\mathrm{orb} \leq 100~\text{days}$. The position of Kepler-432 b is marked 
with a green triangle in both panels.}
\label{fig:3}
\end{figure}

After \object{Kepler-91 b} and \object{HIP67851 b}, Kepler-432 b is the third planet found to orbit a giant star at a 
distance closer than $a=0.5$~AU (Fig. \ref{fig:3}), and the second found to transit a star ascending the RGB.  Although the current 
sample is not statistically significant, the discovery of Kepler-432 b confirms that close-in planets around 
intermediate-mass giant stars do exist. Given the short dissipation time-scale of protoplanetary disks, gravitational interaction seems 
to be the favorite migration channel for close-in planets of intermediate-mass stars. Their paucity might be ascribed to enhanced 
tidal dissipation and subsequent stellar engulfment during the RGB phase.  

\citet{Villaver2009} and \citet{Kunitomo2011} have computed the planetary orbit evolution during the post MS phase for stars more 
massive than the Sun. They concluded that short-period planets are swallowed by their host stars as a result of the increased star-planet 
tidal interaction during the RGB phase. Both studies predict a critical semi-major axis below which planets are engulfed by their 
host star. The expected values for a star as massive as Kepler-432 are much higher than the current semi-major axis of Kepler-432 b. 
This implies that the planet will not survive the RGB phase and will be swallowed by its host star before it reaches the tip of the RGB.

\section{Conclusions}
We spectroscopically confirmed the planetary nature of the transiting candidate Kepler-432 b, derived a planetary mass of 
\mp=$5.84\pm0.05$~\Mjup, and found that the orbit is eccentric ($e$=$0.478\pm0.004$). Kepler-432 b is the first 
\textit{bona fide} warm Jupiter planet found to orbit a giant star, and, after \object{Kepler-91 b}, is the second found to transit a giant star. The semi-major 
axis $a$=$0.303\pm0.007$~AU and eccentricity of the planet suggest that some kind 
of migration mechanism must have operated (or is operating now) to bring the planet to its current position. In this context, we 
discussed the high-eccentricity migration scenario as a plausible mechanism for the formation of this system.  This possibility, 
although not directly verifiable with the currently available data, would account for the high eccentricity and small semi-major 
axis of the planet, provided that a second massive planet is also orbiting the system. The latter scenario is corroborated by the 
detection of a significant radial velocity trend in our data. Additional spectroscopic follow-up observations are needed to better 
characterize the system.

Although our discovery confirms that close-in ($a\lesssim0.5$~AU) giant planets can exist around giant stars, more detections are 
needed to properly characterize the population of these objects around post-MS stars. According to current post-MS evolutionary 
models, Kepler-432 b will not survive the RGB phase and will be engulfed by its host star.

\emph{\noteaddname} This letter was submitted in parallel with that of \cite{Ciceri2014}, who independently also confirmed the planetary 
nature of Kepler-432 b. Their results agree with ours within the observational errors.

\begin{acknowledgements}
      M.O. and S.R. acknowledge funding from the 
      \emph{Deut\-sche For\-schungs\-ge\-mein\-schaft, DFG\/}   under project number RE 2694/3-1 611328 to carry out the observations at 
      Calar Alto. H.J.D and D.N acknowledge support by grant AYA2012-39346-C02-02 of the Spanish Ministerio de Econom\'ia y Competitividad. 
      We are very grateful to the staff members at Calar Alto and at the Nordic Optical Telescope for their valuable and unique support 
      during the observations. M.O. thanks N. Kudryavtseva and S. Ciceri for useful discussion. 
 
\end{acknowledgements}


\Online

\end{document}